\documentclass[twocolumn,aps,prl]{revtex4}%
\usepackage{amsmath}
\usepackage{graphicx}
\usepackage{dcolumn}
\usepackage{bm}%
\usepackage{amsfonts}%
\usepackage{amssymb}
\preprint{}

\begin{document}

\title{Universal Description of Spherical Aberration Free Lenses Composed of Positive
or Negative Index Media}
\author{D. Schurig and D. R. Smith}
\affiliation{Physics Department, University of California, San Diego, La Jolla, CA, 92093}
\date{\today   }

\begin{abstract}
We find that the function that describes the surface of spherical aberration
free lenses can be used for both positive and negative refractive index media.
With the inclusion of negative index, this function assumes the form of all
the conic sections and expands the theory of aplanatic optical surfaces.
\ There are two different symmetry centers with respect to the index that
create an asymmetric relationship between positive and negative index lens
profiles. In the thin lens limit the familiar formulas for image position and
magnification hold for any index.

\end{abstract}
\maketitle

It was known to Ren\'{e} Descartes as early as 1637 that aspherical surfaces
are required to convert plane waves into spherical waves and vice versa. In
the short wavelength limit of geometric optics, the right conic surface can
bring all parallel incident light rays to a single common focus point.\ Most
commercial lenses have spherical profiles and only perform this function
approximately; they are said to have \emph{spherical aberration}. It is only
somewhat well known that a hyperboloid is the ideal lens shape for this
purpose when using media with a relative refractive index greater than one.
Less well known, is the fact that an ellipsoid is the ideal shape for a lens
composed of media with relative index between zero and one \cite{hecht}.

Much of the fundamental behavior of media with negative refractive index was
predicted in 1968 by Veselago \cite{veselago}.\ Such media are not naturally
occurring, so their properties were not extensively pursued at that time.
Recently, negative index media were demonstrated using engineered composites
\cite{smithPRL,shelby}, and the same design paradigm used for those composites
also enables the construction of media with index less than one and even
somewhat close to zero. This makes the question of optical element design with
these media pertinent. In this paper we will discuss ideal (i.e. spherical
aberration free) shapes for lenses composed of media with both positive and
negative index. We will derive universal formulas for lens design that apply
to media of any index, and discuss the symmetry between positive and negative
index lenses. \ We will also demonstrate that, with the inclusion\ of negative
index, a more compete theory of aplanatic optical surfaces is obtained.

The lenses discussed here need to be clearly distinguished from the
``perfect'' lens of Pendry \cite{perfectLens}. These lenses can focus incident
plane waves, i.e. waves from a source object at infinite distance. They are
said to possess an aplanatic point at infinity \cite{bornWolf}. Plane waves
are \emph{not} focused by the ``perfect'' lens; it has aplanatic points only
at distances within one lens thickness. The importance of the ``perfect'' lens
lies in its near field focusing capability. This work describes lenses that
can focus or collect energy from distant sources where the lenses are composed
of materials from an extended parameter space, namely negative refractive
index.%
\begin{figure}
[t]
\begin{center}
\includegraphics[
height=2.1205in,
width=2.9603in
]%
{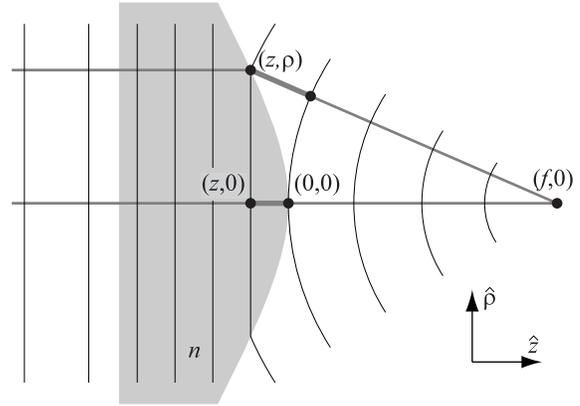}%
\caption{Optical paths for the central and a peripheral ray. \ In the
transition region wave fronts intersect the lens surface and lie both inside
and outside lens. \ The two thicker segments in the transition region must
have equal optical path length. \ The point $(z,r)$ lies on the desired
surface.}%
\label{opl fig}%
\end{center}
\end{figure}

Aspherical lenses are uncommon because they are difficult to manufacture.\ The
requirement of $\lambda/4$ to $\lambda/10$ surface roughness for good optics
is quite difficult to achieve in the visible range for non spherical surfaces.
However, this requirement is automatically satisfied for surfaces on the
composite materials that have been used to implement negative index media.
These media are composed of unit cells that must be significantly smaller than
the operational wavelength in order for them to function as a homogenous media
\cite{xlowFreqPlasmons,smithPRL}.\ The manufacturing process that enables
construction on the required sub unit cell length scales can by default
control the surface profile with one unit cell accuracy. Though the current
technology may never extend to visible light, whatever the operational
wavelength of these media, sub-wavelength surface accuracy for any desired
shape is available.

The results discussed here were first obtained by ray tracing \cite{hecht}.
The ray tracing algorithm employed only fundamental assumptions about the
media dispersion , boundary matching, and the conservation of energy. For
isotropic media, our algorithm is equivalent to Snell's law, which applies to
interfaces with relative index of either sign
\cite{shelby,parazzoliSnell,pachecoDispersion,FoteinopoulouFDTD}. The
conservation of energy was ensured using the Poynting vector, with each ray
carrying power into a surface being required to carry power out.%
\begin{figure}
[t]
\begin{center}
\includegraphics[
height=5.6377in,
width=3.4601in
]%
{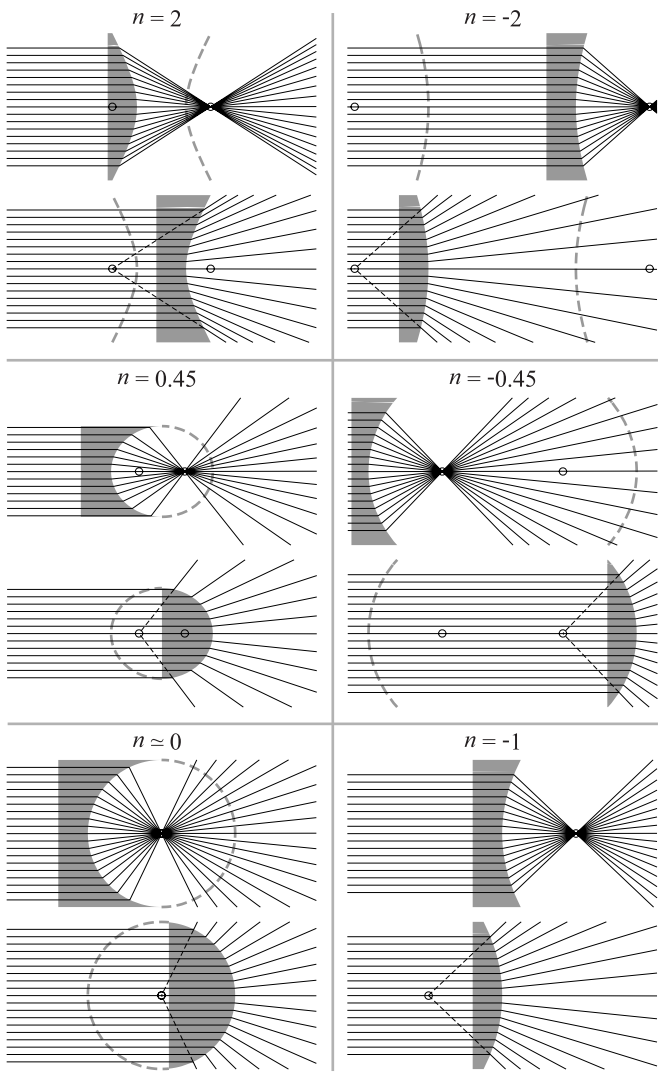}%
\caption{Spherical aberration free converging and diverging lenses composed of
positive and negative refracting media. \ The unused portion of the conic
section is shown dashed. \ Foci are shown as circles \ Virtual image rays are
shown dashed. All lenses have the same focal length.}%
\label{ray fig}%
\end{center}
\end{figure}

Here, however, we show an analytical argument using optical path length (OPL)
following Hecht \cite{hecht}. We wish to find a surface that refracts parallel
rays to a single point (Fig. \ref{opl fig}). Parallel rays have planar wave
fronts on which the phase is constant. Rays converging to a point have
spherical wave fronts on which the phase is constant. The desired lens surface
must be contoured such that, the optical path length in the transition region
is the same for all rays. The optical path length or phase advance
(retardation) is \cite{hecht, bornWolf}
\begin{equation}
OPL=\int_{C}nds\label{opl eq}%
\end{equation}
The equivalence of the \emph{OPL} in the transition region for the two rays
shown in Fig.\ref{opl fig} yields,%
\begin{equation}
\sqrt{\left(  f-z\right)  ^{2}+\rho^{2}}-f=-nz
\end{equation}
From this we find%
\begin{equation}
\frac{\left(  z-a\right)  ^{2}}{a^{2}}+\frac{\rho^{2}}{b^{2}}%
=1\label{conic eq}%
\end{equation}
which is an equation for conic sections where the $a,b$ parameters are given
in terms of the index, $n$, and focal distance, $f$.
\begin{subequations}
\label{conic params eq}%
\begin{align}
a  &  =f\frac{1}{1+n}\\
b^{2}  &  =f^{2}\frac{1-n}{1+n}%
\end{align}
Since $b^{2}$ can be negative, Eq. (\ref{conic eq}) can and does generate all
the conic sections for different values of the index. Further, nothing in this
derivation precludes $n$ from being negative, as long as the optical path
length as given by Eq. (\ref{opl eq}) is applicable. We can see this as
follows. Geometric rays follow the energy direction indicated by the Poynting
vector. It is well established that, in isotropic negative index media, the
phase advance direction is opposite to the Poynting vector
\cite{veselago,modulate,pachecoDispersion,FoteinopoulouFDTD}, so it is
appropriate to use negative index in Eq. (\ref{opl eq}).

The focal length, $f$, can also be negative to find lens surfaces that
generate diverging spherical waves from plane waves (i.e. a virtual focus).
The following functional form gives a surface centered at $z=r=0$ with the
correct concavity for plane waves on the negative $z$ side and spherical waves
on the positive $z$ side.
\end{subequations}
\begin{equation}
z(\rho)=a\left(  1-\sqrt{1-\frac{\rho^{2}}{b^{2}}}\right) \label{function eq}%
\end{equation}
This equation is valid for all values of $f$ and $n$, except $n=-1$, where it
is singular. This limiting case is handled by letting $n=-1+\delta$ with
$\delta\ll1$, and substituting into Eq. (\ref{function eq}) with Eq.
(\ref{conic params eq}). We then find%
\begin{equation}
z(\rho)\simeq\frac{1}{4f}\rho^{2}\label{parabaloid eq}%
\end{equation}
which is a paraboloid.

From Eqs. (\ref{conic params eq}),(\ref{function eq}) and (\ref{parabaloid
eq}), we find the following behavior for the ideal lens surface. These results
are also confirmed by ray tracing, Fig. \ref{ray fig}.
\[%
\begin{tabular}
[c]{|c|c|c|c|}\hline
\textbf{index} & \textbf{conic} & \textbf{converge/diverge} & \textbf{focus}%
\\\hline\hline
$1<n$ & hyperboloid & convex/concave & far\\\hline
$n=1$ & no solution & - & -\\\hline
$0<n<1$ & ellipsoid & concave/convex & far\\\hline
$n=0$ & sphere & concave/convex & far = near\\\hline
$-1<n<0$ & ellipsoid & concave/convex & near\\\hline
$n=-1$ & paraboloid & concave/convex & near\\\hline
$n<-1$ & hyperboloid & concave/convex & near\\\hline
\end{tabular}
\]%
\begin{figure}
[t]
\begin{center}
\includegraphics[
height=3.4592in,
width=3.2603in
]%
{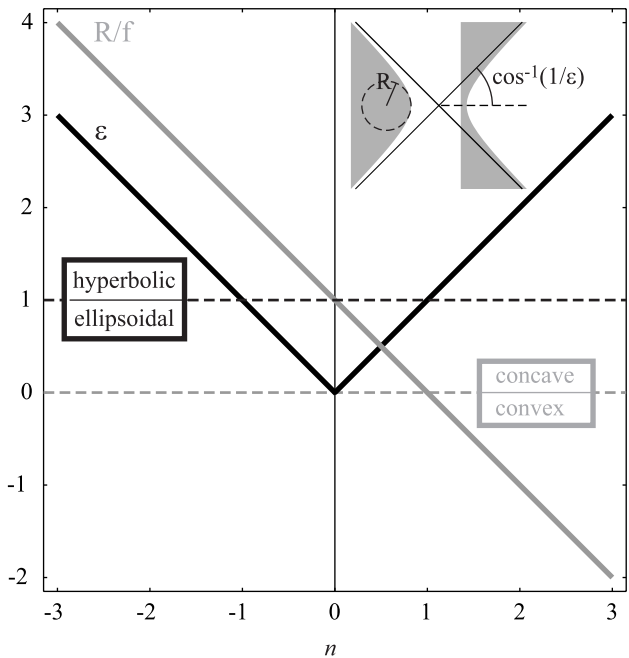}%
\caption{Eccentricity (black line) and major axis radius of curvature (gray
line) vs. refracive index. An eccentricity, $\varepsilon$, of one is the
boundary between hyperbolic and ellisoidal lens surfaces. \ A radius, $R$, of
zero is the boundary between concave and convex surfaces. \ The concavity
indicated on this boundary in the figure is for converging lenses. \ The inset
shows the definition of $R$ and $\varepsilon$ for concave and covex hyperbolic
lenses.}%
\label{graph fig}%
\end{center}
\end{figure}

The conic sections characterized by the $a,b$ parameters above, can be equally
well characterized by another set of parameters, the eccentricity,
$\varepsilon$ and the major axis radius of curvature, $R$, (Fig. \ref{graph
fig} inset). This pair of parameters is useful in that it separates the shape,
given by $\varepsilon,$ from the length scale, given by $R$. Using the usual
definition, eccentricities less than one indicate ellipses and greater than
one indicate hyperbolas. Applying Eq. (\ref{conic params eq}) leads to the
simplification%
\begin{equation}
\varepsilon\equiv\frac{\sqrt{a^{2}-b^{2}}}{\left|  a\right|  }=\left|
n\right|
\end{equation}
Thus the eccentricity is symmetric in $n$, (Fig. \ref{graph fig}). If this
were the only parameter, we could take any positive index lens design and use
it for lenses composed of negative index media.

The radius of curvature can be calculated from Eq. (\ref{function eq}). \
\begin{equation}
R\equiv\frac{1}{z^{\prime\prime}\left(  0\right)  }=\frac{b^{2}}{a}=f(1-n)
\end{equation}
This is recognized as the familiar lens makers formula usually derived using
spherical optics with the paraxial approximation. This is not surprising since
any of the conic sections approximate a sphere when examined with small enough
aperture; the two derivations must agree on the lens central radius of
curvature. Ray tracing confirms this formula applies for both positive and
negative index and correctly gives a change in the concavity at $n=1$. Because
this formula is antisymmetric about $n=1$, (Fig. \ref{graph fig}), we cannot
use the same lens design for positive and negative index. If we do, the
positive and negative index lenses will have different focal lengths. In fact,
for $\left|  n\right|  >1$, one lens will be converging and the other
diverging. Notably, the positive index analog of the $n=-1$ parabolic lens is
absent, since the $n=1$ lens requires zero radius of curvature.

We have not yet shown how, if at all, the position of the optical focus of one
of these lenses relates to the geometric foci of its conic surface. It can be
easily shown algebraically or by ray tracing, that the optical focus lies
exactly at the position of one of the conic section's foci. Ray tracing also
confirms a simple rule that specifies which focus. For positive index the
correct focus is the one furthest from the branch of the conic that is used
for the lens surface. For negative index, the opposite is true. For the lenses
with parabolic ($n=-1$) and spherical ($n=0$) surfaces that posses just one
geometric focus, it is also the optical focus.%

\begin{figure}
[t]
\begin{center}
\includegraphics[
height=1.7729in,
width=3.4307in
]%
{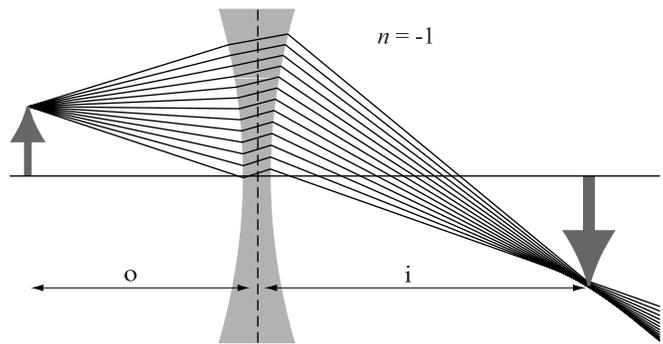}%
\caption{Magnification by a spherical aberration free, bi-concave lens
composed of $n=-1$ media.}%
\label{magray fig}%
\end{center}
\end{figure}
As pointed out in Hecht \cite{hecht}, two conic surfaces can be combined to
construct lenses that convert diverging spherical waves to converging
spherical waves. Ray tracing confirms that in the thin lens limit, the usual
formulas apply for image distance%
\begin{equation}
\frac{1}{f}=\frac{1}{o}+\frac{1}{i}%
\end{equation}
and magnification%
\begin{equation}
M=-\frac{i}{o}=\frac{f}{f-o}%
\end{equation}
regardless of the sign of the index. In Fig. \ref{magray fig}, a double
concave parabolic lens is shown to function as a converging lens when composed
of $n=-1$ material.

The extension of spherical aberration free lens design to negative index media
yields a more complete theory, but the practical limitations of current
negative index media, notably strong chromatic aberration (due to media
dispersion) may preclude technological interest in frequency ranges where good
positive index materials are available. However, in frequency ranges where
this is not the case, (e.g. the millimeter wave range), and composite media
are already in use, negative index lenses can be advantageous. Where larger
radius of curvature is desirable, it is worth noting that an $n=-2$ lens has
three times the radius of curvature of an $n=2 $ lens of the same focal
length, (Fig. \ref{ray fig}). Even more significant is the fact that an $n=-1$
lens is impedance matched to the surrounding media \cite{perfectLens};
\emph{none} of the incident radiation is reflected. This is impossible with a
positive index lens.

\begin{acknowledgments}
This work was supported by DARPA Contract No. MDA972-01-2-0016 and by DARPA
through a grant from ONR, Contract No. N00014-00-1-0632.
\end{acknowledgments}

\bibliographystyle{prsty}
\bibliography{dav}
\end{document}